\documentclass[notitlepage,a4paper,aps,prd,twocolumn,superscriptaddress,nofootinbib,groupedaddress]{revtex4}
\usepackage{graphicx}
\usepackage[colorlinks=true, pdfstartview=FitV, linkcolor=blue, citecolor=red, urlcolor=magenta]{hyperref}

\usepackage{graphics}
\usepackage{amssymb}
\usepackage{amsmath}
\usepackage{url}
\usepackage{color}
\usepackage[utf8]{inputenc}
\usepackage{ulem}
\usepackage{tikz}
\usetikzlibrary{arrows,decorations.markings}

\newcommand{\fracc}[2]{\frac{\textstyle{#1}}{\textstyle{#2}}}

\newcommand{\x}{\vec{x}}
\newcommand{\rmd}{\textrm{d}}
\newcommand{\e}{\textrm{e}}
\begin{document}
\newcommand{\Areia}{
\affiliation{Department of Chemistry and Physics, Federal University of Para\'iba, Rodovia BR 079 - Km 12, 58397-000 Areia-PB,  Brazil}
}
\newcommand{\Lavras}{
\affiliation{Physics Department, Federal University of Lavras, Caixa Postal 3037, 37200-000 Lavras-MG, Brazil}
}
\newcommand{\unifei}{\affiliation{Federal University of Itajub\' a, Av.\ BPS 1303, Itajub\' a-MG, 37500-903, Brazil}}

\newcommand{\ufes}{\affiliation{PPGCosmo, CCE - Federal University of Esp\'{\i}rito Santo, Vit\'{o}ria-ES, 29075-910, Brazil.}}

\title{Interference effects and modified Born rule in the presence of torsion}

\author{Eduardo Bittencourt}\email{bittencourt@unifei.edu.br}
\unifei
\author{Alexsandre L. Ferreira Junior}\email{alexsandre.ferreira@edu.ufes.br}
\unifei
\ufes
\author{Iarley P. Lobo}\email{iarley\_lobo@fisica.ufpb.br}
\Areia
\Lavras

\date{\today}

\begin{abstract}
The propagation of nonrelativistic excitations in material media with topological defects can be modeled in terms of an external torsion field modifying the Schr\"odinger equation. Through a perturbative approach, we find a solution for the wave function which gives corrections in the interference patterns of the order of $0.1\, \textrm{\AA}$, for a possible experimental setup at atomic scales. Finally, we demonstrate how this geometric, but effective, approach can indeed accommodate a probabilistic interpretation of the wave function although the perturbative theory is nonunitary.
\end{abstract}

\maketitle

\section{Introduction}
Generalizations of the spacetime structure concerning the relationship between the metric and the affine connection have been pursued since after the proposal of general relativity \cite{weyl1}. In the last decades, the extra degrees of freedom of the connection, beyond the Levi-Civita one, have been used to produce cosmic inflation \cite{Ghilencea:2019rqj}, explain dark matter \cite{Israelit:1992ae} or model CPT violation \cite{Lobo:2018zrz,Delhom:2020vpe}. The skew-symmetric part of the connection, called \textit{torsion tensor}, may play an important role in the description of spacetime \cite{BeltranJimenez:2019tjy}, being a useful tool for alternative theories of gravity \cite{Krssak:2015oua, cabral2020,Bahamonde:2020vpb}, super-symmetry or quantum gravity \cite{Cabral:2020fax}. There are many different models found in the literature dealing with this field along one of the following approaches (see Ref.\ \cite{flanagan} and references therein): (i) algebraic torsion, (ii) dynamical torsion or (iii) field dependent torsion. On the other hand, experimental tests of gravity have set very narrowed bounds on the possibility of a fundamental torsion field in the spacetime. It means that if this field is necessary for describing the gravitational phenomenon, then the precision of the measurements must be considerably enhanced in order to detect its tiny contribution. That is the conclusion from experiments and simulations like LAGEOS, Gravity Probe B and BepiColombo \cite{lageos,guth,gravprobeB,schettino}. 

At the quantum level, the influence of torsion can be mathematically accommodated assuming the principle of general covariance. From the physical viewpoint, spinning or spinless relativistic particles follow well-behaved equations of motion in the presence of torsion, while nonrelativistic spinless particles cannot couple to this field in order that the theory be unitary and Hermitian (for details, see full discussion provided by Ref.\ \cite{kleinert2000}). It is also known that in most theories with torsion, the coupling with nonrelativistic spinless particles does not occur because it is produced by fermionic matter, thus being completely skew-symmetric. However, it has been proposed recently different sources for the torsion field avoiding the previous argument, as one can see in Ref.\ \cite{cabral2020}.

In between, at mesoscopic scales, one can find geometric models taken torsion into account to describe effectively topological defects \cite{kleinert89,katanaev, delorenci,ciappina,imaki2019,randono2011}, like dislocations, and transport properties \cite{gromov2015}  within material media. This is the line we shall follow here, considering a nonrelativistic particle satisfying a modified Schr\"odinger equation in presence of an external pointlike torsion, and presenting a perturbative analytic solution for the wave function, from which we can see how the presence of torsion changes the interference patterns in a possible empirical setting. Such effective approach may shed some light to the role played by torsion in making the theory nonunitary, but with a quasiprobabilistic interpretation in virtue of a measurable distinction from the torsion-free case. With this in mind, we propose a new expression for the inner product in the case of the nongradient torsion and find physical solutions for the wave function, overcoming the difficulties found by the previous studies on this issue.

\section{Schr\"odinger equation in general metric-affine spacetimes}

Following Ref.\ \cite{kleinert2000}, the study of quantum mechanics in an arbitrary metric-affine spacetime can be done by introducing a nonholonomic map, so that
\begin{equation}
dx'^a=e^a_{\hphantom a\mu}(x^\lambda)dx^\mu,
\label{coordtransf}
\end{equation}
where $e^a_{\hphantom a\mu}=\partial x'^a/\partial x^\mu$. Here Greek and lower Latin indexes run from 0 to 3, while capital Latin indexes vary from 1 to 3; exception for $i,j,k$ which also run from 1 to 3. The functions $x'^a(x^\lambda)$ should be multivalued in order to take into account curvature and torsion. Thus, the functions $e^a_{\hphantom a\mu}(x^\lambda)$ will be a multivalued tetrad basis, yielding the induced metric in the space %spawned
spanned by $x^\lambda$, as follows
\begin{equation}
	g_{\mu\nu}=e^a_{\hphantom a\mu} e^b_{\hphantom a\nu} \eta_{ab}.
	\label{metricdef}
\end{equation}
The reciprocal basis is defined as $
e^{\hphantom a \mu}_{a}\equiv\eta_{ab}g^{\mu\nu}e^b_{\hphantom b\nu}$ and both bases satisfy the identities $
e^a_{\hphantom a\mu}e^{\hphantom a \mu}_{b}=\delta^a_{\hphantom a b}$ and $e^a_{\hphantom a\mu}e^{\hphantom a \nu}_{a}=\delta^{\nu}_{\hphantom a\mu}$, where $\delta^a{}_b$ and $\delta^{\nu}{}_\mu$ are Kronecker deltas. Thence, from $e^a_{\hphantom a \mu}(x^\lambda)$, we construct the connection
\begin{equation}
\Gamma^{\mu}_{\hphantom a\lambda\kappa}=e^{\hphantom a \mu}_{a}\partial_\lambda e^a_{\hphantom a\kappa},
\label{connectiondef}
\end{equation}
from which we define the covariant derivative operator as $D_\mu v^{\alpha} =\partial_\mu v^{\alpha} +\Gamma_{\hphantom a\mu\lambda}^{\alpha}v^{\lambda}$, where $v^{\alpha}$ is an arbitrary vector field. This covariant derivative is such that the metricity condition holds, i.e., $D_\mu g_{\nu\lambda}=0$. The torsion tensor $T^{\lambda}_{\hphantom a\mu\nu}$ in this case is
\begin{equation}
T^{\lambda}_{\hphantom a\mu\nu}=\frac{1}{2}\left(\Gamma^{\lambda}_{\hphantom a\mu\nu}-\Gamma^{\lambda}_{\hphantom a\nu\mu}\right)=\frac{1}{2}e^{\hphantom a \lambda}_a\left(\partial_\mu e^a_{\hphantom a \nu}-\partial_\nu e^a_{\hphantom a \mu}\right),
\label{tensordef}
\end{equation}
while the curvature tensor is defined from the connection as
\begin{equation}
R^\lambda_{\hphantom a\mu\nu\kappa}=e^{\hphantom a \lambda}_a\left(\partial_\mu\partial_\nu-\partial_\nu\partial_\mu\right)e^a_{\hphantom a \kappa}.
\label{curvaturedef}
\end{equation}
Note that for nonvanishing torsion and curvature, the Schwarz integrability condition should not be satisfied by the coordinates $x'^a(x^\lambda)$ and by the basis $e^a_{\hphantom a \mu}(x^\lambda)$, respectively. This is crucial to distinguish our multivalued basis from the usual vierbein, defined in the theories of spinning particles in curved spacetimes. The latter is single-valued, taking every point of the manifold into Minkowski, therefore carrying no curvature or torsion. While the former maps a small neighborhood of the manifold into a flat spacetime. The relation between them is given by a Lorentz transformation, which must be multivalued (see Sec. 13 of Ref.\ \cite{kleinert2000} for further discussion).

In order to describe the quantum dynamics of a nonrelativistic system in this general metric-affine space, we shall use the minimal coupling prescription to substitute the partial derivative for the covariant derivative. In this way, the Schr\"{o}dinger equation becomes
\begin{eqnarray}
i\hbar \frac{\partial}{\partial t} \Psi (x^\lambda)=-\frac{\hbar^2}{2m}D_i D^i \Psi (x^\lambda)+V(x^\lambda)\Psi (x^\lambda)\nonumber\\
=\frac{\hbar^2}{2m}\left(\partial_i \partial^i+\Gamma^{i}_{\hphantom a i j}\partial^{j}\right) \Psi (x^\lambda)+V(x^\lambda)\Psi (x^\lambda),
 \label{schroedingereq1}
\end{eqnarray}
where being time an absolute quantity, only spatial torsion and curvature are presented. The connection term may be split into a symmetric part that defines the unitary Laplace-Beltrami operator together with the Laplacian $\partial_i\partial^i$, while the skew-symmetric term (torsion) is responsible for the nonunitary contribution to the Schr\"odinger equation. It should be noticed that Eq.\ (\ref{schroedingereq1}) is the same one found by Kleinert through a path integral approach \cite{kleinert2000}.

Finally, for the introduction of a time component in the torsion, we must transform the derivatives by the nonholonomical mapping $\partial_a\rightarrow e^{\hphantom a\mu}_a\partial_\mu$, which leads to
\begin{eqnarray}
 i\hbar e^{\hphantom a\mu}_0\partial_\mu\Psi =-\frac{\hbar^2}{2m}\left(g^{\mu\nu}\partial_\mu \partial_\nu+\Gamma^{\mu}_{\hphantom a \mu \nu}\partial^\nu\right) \Psi \nonumber\\
 +\frac{\hbar^2}{2m}\left[e^{\hphantom a\mu}_0 e^{ 0}_{\hphantom a\nu}\partial_\mu \partial^\nu+e^{\hphantom a\mu}_0\left(\partial_\mu  e^{ 0}_{\hphantom a\nu}\right)\partial^\nu\right] \Psi +V\,\Psi.
 \label{schroedingereq2}
\end{eqnarray}
Note that we recover the absolute time expression when $e^{\hphantom a 0}_0=1$ and $e^{\hphantom a i}_0=e^0_{\hphantom a i}=e^{\hphantom a 0}_A=e^A_{\hphantom a 0}=0$.
It is straightforward to see that Eq.\ (\ref{schroedingereq2}) has a immediate correspondence to the Schr\"{o}dinger equation written in a Newton-Cartan background \cite{gromov2015}. 

\section{Torsion generated by a screw dislocation}

In this vein, we shall investigate the motion of nonrelativistic particles in the particular case of a two-dimensional crystalline lattice with a screw dislocation in a given spatial direction. Usually, such defect in a medium is attributed to the removal of layers of atoms from the lattice. A partial description of this system, where the defect is modeled by torsion can be found in \cite{kleinert2000}. However, we shall demonstrate that it is possible to go further and bypass the complaints encountered in the full theory, as non-Hermiticity and nonunitarity, assuming an effective description of the system where the conservation of probability can be neglected at a first glance.

In our case, we can map the distorted coordinates $x'^{i}=(x'^{1},x'^{2})$ of the atoms in the lattice to the undistorted ones $x^{i}=(x^{1},x^{2})$ through the parametrization
\begin{equation}
    dx'^{1}=dx^1, \hspace{5mm} dx'^2=dx^2+\varepsilon\partial_i\phi(x^j)dx^i,
\end{equation}
with $\varepsilon=nd \ll 1$, where $n$ is the number of planes removed from the crystalline lattice and $d$ is the distance between two adjacent planes, so that terms of order $\varepsilon^2$ are negligible. Latin indexes shall assume the values $1$ and $2$ from now on. The function $\phi(x^j)$ is the multivalued function
\begin{equation}
\phi(x^j)=\arctan\left(\frac{x^2}{x^1}\right),
\end{equation}
that is made unique by cutting the plane in a line from the origin to infinity, and demanding the function to go from $\pi$ to $-\pi$ when passing through it.

Now, let us consider a small closed path $C$ encircling the origin, and the surface $S$ enclosed by it. The integration of the commutator of partial derivatives of $\phi(x^j)$ over $S$ yields
\begin{equation}
  \int_S d^2x (\partial_1\partial_2-\partial_2\partial_1)\phi(x^j)=\oint_C dx^i \partial_i \phi(x^j)=2\pi. \end{equation}
Further, our multivalued basis is given by
\begin{equation}
    e^1_{\hphantom a i}=\delta^1_i, \hspace{5mm} e^2_{\hphantom a i}=\delta^2_i+\varepsilon\partial_i\phi(x^j),
\end{equation}
and the induced metric, given by Eq.\ (\ref{metricdef}), reads
\begin{equation}
    g=\begin{pmatrix}
1 & \varepsilon\partial_1\phi(x^j)\\
\varepsilon\partial_1\phi(x^j) & 1+2\varepsilon\partial_2\phi(x^j)
\end{pmatrix}.
\end{equation}
From this, we find the inverse tetrad basis as
\begin{equation}
    e_1^{\hphantom a i}=\delta_1^i-\delta_2^i\varepsilon\partial_1\phi(x^j), \hspace{5mm} e_2^{\hphantom a i}=\delta_2^i\big[1-\varepsilon\partial_i\phi(x^j)\big].
\end{equation}
Since the derivative of $\phi(x^j)$ is a single-valued function, the curvature tensor is zero. The connection coefficients read
\begin{equation}
    \Gamma^{i}_{\hphantom a jk}=e_A^{\hphantom a i}\partial_j e^A_{\hphantom a k}=
    e_2^{\hphantom a i}\partial_j e^2_{\hphantom a k}=\delta^i_2\varepsilon\partial_j\partial_k\phi(x^l),
\end{equation}
and finally, the torsion tensor becomes 
\begin{equation}
    T^{1}_{\hphantom a ij}=0,\hspace{5mm} T^{2}_{\hphantom a ij}=\frac{\varepsilon}{2\pi}(\partial_i\partial_j-\partial_j\partial_i)\phi(\x)=\epsilon_{ij}\varepsilon\delta(\x).
    \end{equation}
One can also introduce a discrete numbers $n$ of defects at different points $(x^1_n,x^2_n)$ of space by choosing the multivalued function $\phi(x^j)$ to be
\begin{equation}
\phi(x^j)=\sum_n \mathrm{arctan}\left(\frac{x^2-x^2_n}{x^1-x^1_n}\right).
\label{moredef}
\end{equation}
Thus, the corresponding torsion field will be the superposition of the pointlike torsions as
\begin{equation}
T^{1}_{\hphantom a ij}=0,\hspace{5mm} T^{2}_{\hphantom a ij}=\epsilon_{ij}\varepsilon\sum_n \delta(\x-\x_n).
\end{equation}
However, we cannot have a continuous distribution of defects because the endpoints of the branch cuts must remain distinguishable \cite{kleinert2000}.
    
\section{Carrying out the torsion in the Schr\"{o}dinger equation}
In virtue of the nonunitarity problem of the Schr\"odinger equation in the presence of torsion, the previous studies do not investigate possible solutions for Eq.\ (\ref{schroedingereq1}). In fact, this has been a disadvantage against further analysis of this equation whose the situation is typically circumvented by claiming that the torsion field can be implemented only within the relativistic quantum mechanics due to the nature of the torsion generated by matter fields (see Ref.\ \cite{kleinert2000} and references therein).

\begin{figure}[!ht]
\includegraphics[width=0.37\textwidth]{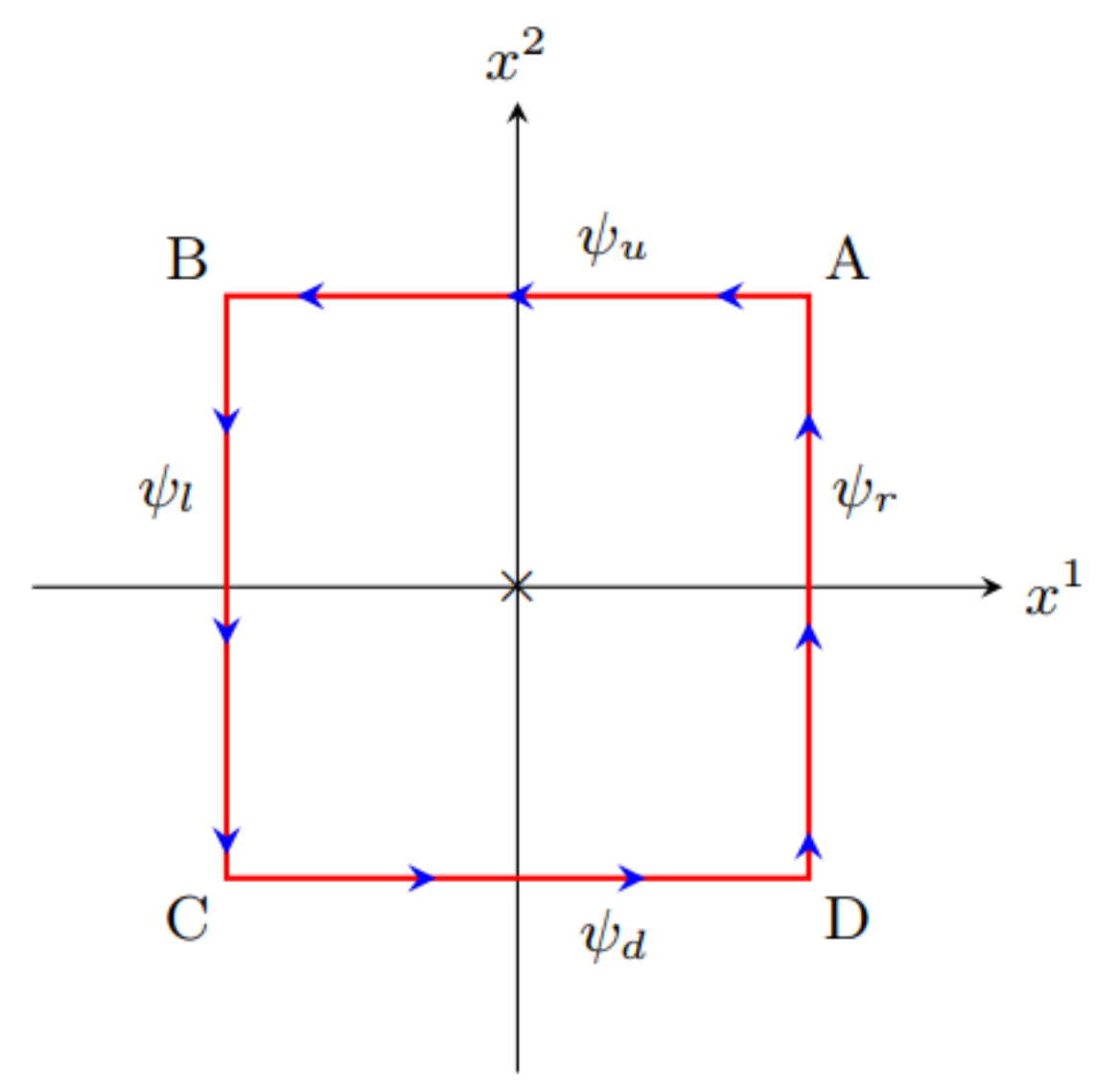}
\caption{Oriented squared region of side $\eta$ enclosing the origin in the $(x^1,x^2)$-plane.}
\label{fig:1}
\end{figure}

From the previous sections, we were able to implement torsion effects in the Schr\"{o}dinger equation through modifications in the Laplace operator due to the non-Riemannian connection and discuss the use of the torsion field to describe topological defects in crystalline lattices. Thus, Eq.\ (\ref{schroedingereq1}) for a free particle moving in such crystalline surface becomes
\begin{equation}
i\hbar\fracc{\partial}{\partial t} \Psi = -\fracc{\hbar^2}{2m} \left(\partial_i\partial^i + (\partial_i\Lambda)\partial^i - T^{2}_{\hphantom a 12}\partial_1\right)\Psi,
\end{equation}
with $\Lambda=\varepsilon\partial_2\phi(x^i)$. Substituting $\Psi (x^i,t)= \textrm{e}^{-\frac{\Lambda}{2}}\psi (x^i,t)$ in the equation above and neglecting terms quadratic in $\varepsilon$, we find that
\begin{equation}
    i\frac{\partial}{\partial t}\psi =-\frac{\hbar}{2m}\left(\partial_i\partial^i\psi -T^{2}_{\hphantom a 12}\partial_1\psi -\frac{1}{2}\psi\, \partial_i\partial^i\Lambda\right)\, .
\end{equation}
Straightforward calculations show that the potential-like term vanishes, that is, $\partial_i\partial^i\Lambda=0$. Thence,
\begin{equation}
    \frac{2i m}{\hbar} \frac{\partial}{\partial t}\psi+\partial_i\partial^i\psi =-\varepsilon\delta(x^i)\partial_1\psi.
    \label{eq30}
\end{equation}
Now, we shall solve this equation perturbatively in powers of $\varepsilon$. Thus, suppose a solution of the form $\psi(x^i,t)=\psi_0(x^i,t)+\varepsilon\psi_1(x^i,t)$, where $\psi_0$ is the solution to the homogeneous equation away from the origin. So, $\psi_1$ must satisfy
\begin{equation}
\label{dif_eq_psi1}
     \frac{2i m}{\hbar} \frac{\partial}{\partial t}\psi_1+\nabla^2\psi_1 =-\delta(x^i)\partial_1\psi_0\, .
\end{equation}
Integrating this equation over a small region $S$ enclosing the origin with the use of the divergence theorem in two dimensions \cite{butkov1973}, yields
\begin{equation}
\label{int_eq_psi1}
 \frac{2i m}{\hbar} \frac{\partial}{\partial t} \int_S \textrm{d}^2x\, \psi_1+\oint_{C}\textrm{d}s \,n^i\partial_i\psi_1=- \partial_1\psi_0\big|_{x^i=0}\, ,
\end{equation}
where $s$ is the arc length of $C$ and $n^i$ corresponds to the components of the outward pointing unit normal vector on the boundary of $S$. Notice that we can approximate $\sqrt{g(x^i)}\approx1$, because $\psi_1$ already contributes with a first order term in $\varepsilon$. Moreover, any arbitrary path that encloses the origin can be decomposed into a region that does not contain the origin and, for simplicity, a squared one of small side $\eta\ll1$ surrounding the origin, as depicted in Fig.\ (\ref{fig:1}).

For the region that does not contain the origin, the right-hand side of Eq.\ (\ref{dif_eq_psi1}) is zero, and $\psi$ behaves as the free particle wave function. Nonetheless, for the small squared region in the limit $\eta\rightarrow 0$, the first integral on the left-hand side of Eq.\ (\ref{int_eq_psi1}) can be neglected, while the second term remains:
\begin{eqnarray}
\lim_{\eta\rightarrow0}\left[\int_{-\eta/2}^{\eta/2}\textrm{d}x^1\partial_2\left(\psi_u-\psi_d\right)\right.\nonumber\\
\left.+\int_{-\eta/2}^{\eta/2}\textrm{d}x^2\partial_1\left(\psi_r-\psi_l\right)\right]=- \partial_1\psi_0\big|_{\x=0},
\label{eq31}
\end{eqnarray}
where $\psi_u, \psi_d$ evaluate the wave function along the horizontal lines and $\psi_r, \psi_l$ are the branches of the wave function along the vertical lines (see Fig.\ \ref{fig:1}). From this, the only possibility for $\psi_1$ be continuous at the origin is whether $\partial_1\psi_0\big|_{\x=0}=0$. Otherwise, the first order correction of the wave function has a discontinuity at least in one of its branches.

In what follows, we shall apply this technique to further analyze the consequences of a nonzero torsion field in the Schr\"odinger equation. In particular, as the effects of torsion are manifest only in closed paths, we shall consider interference phenomena through the double-slit experiment for both the cases of plane waves and wave packets.

\subsection{Plane wave solution}
In an interference experiment, such as the double-slit, it is supposed an infinite plate placed at $x^1=0$ with two slits equidistant from the origin, without loss of generality. The state $\psi_0(x^i)=A\textrm{e}^{ik_jx^j}$ is prepared somewhere before the plate, where $x^1<0$ and $x^2=0$, then it is divided into two branches, which we shall call \textit{right-going} $\psi_R(x^i)$ and \textit{left-going} $\psi_L(x^i)$ wave, as it passes through the slits. Setting this apparatus in the lattice, we can put over the plate the semiline generated by the screw dislocation as starting from the origin. As long as the wave functions pass through this dislocation and present a discontinuity, it does not matter which branch it is since the defect only accounts for closed paths. Here, we will choose the right-going wave to stay the same, $\psi_R=\psi_0$, while the left-going one is modified due to the effective torsion, according to
\begin{equation}
\lim_{x^1\rightarrow0^+}\psi_{L}(x^i,t) - \lim_{x^1\rightarrow0^-}\psi_{L}(x^i,t)= \varepsilon A\,B_L\, .
\label{discon}
\end{equation}
Then, from Eq.\ (\ref{eq31}), we find that $B_L=-i\,k^2_1k_2/k^2$, where $k^2=k_1^2+k_2^2$. So, after crossing the defect (region $x^1>0$), the left-going wave is
\begin{equation}
    \psi_L(x^i)=A(1+\varepsilon B_L)\textrm{e}^{ik_jx^j}=A\textrm{e}^{i(k_jx^j-\theta)}+\mathcal{O}(\varepsilon^2)\, ,
\end{equation}
with $\theta=\arctan(\varepsilon k^2_1k_2/k^2)$.

From the usual definition of the probability density
\begin{eqnarray}\label{prob-dens}
\mathcal{P}(x^i,t)\sqrt{g(x^i)}\rmd^2x=|\Psi(x^i,t)|^2\sqrt{g(x^i)}\rmd^2x\nonumber\\
= \textrm{e}^{-\Lambda(x^i)}|\psi(x^i,t)|^2\sqrt{g(x^i)}\rmd^2x\approx|\psi(x^i,t)|^2\rmd^2x,
\end{eqnarray}
with $\e^{-\Lambda}\sqrt{g(x^i)}\approx(1-\Lambda)(1+\Lambda)\approx1$ at first order in $\varepsilon$, one would expect to measure on the screen the quantities
\begin{eqnarray}
    |\psi|^2=|\psi_R(x^i_{(1)})+\psi_L(x^i_{(2)})|^2\nonumber\\
    =4A^2\textrm{cos}^2\left[\frac{k_i(x^i_{(2)}-x^i_{(1)})}{2}-\frac{\theta}{2}\right]\, ,
\end{eqnarray}
where $x^i_{(1)}$ and $x^i_{(2)}$ denote the distances from the slits on the right and on the left, respectively. It is easy see that $|\psi|^2$ would be the same if we have chosen the other way round as the discontinuous branch. Note that this is similar to the well-known electromagnetic Aharonov-Bohm effect, as it introduces a phase shift, although the shift here is proportional to the wave number $k_j$. Such detail is a motivation to analyze the case of wave packets which are quite sensitive to wave numbers. Another interesting point is that if more defects are present through the use of the multivalued function defined in Eq.\ (\ref{moredef}), the modified Schr\"odinger equation becomes
\begin{equation}
    \frac{2i m}{\hbar} \frac{\partial}{\partial t}\psi+\partial_i\partial^i\psi =-\varepsilon\sum_n\delta(x^i-x^i_n)\partial_1\psi,
\end{equation}
which can be solved through the same procedure as before. We single out small squares around each defect enclosed by the wave function, which in turn is cut into left-going and right-going branches in the vicinity of the defect, generating a discontinuity as described by Eq.\ (\ref{discon}). Assuming that the path encloses $q<n$ defects, the total discontinuity will be
\begin{equation}
\sum_{\forall\, \vec p_{q}}\left(\lim_{\vec x\rightarrow \vec{p}_{q}^{\,+}}\psi_{L}(x^i,t) - \lim_{\vec x\rightarrow\vec{p}_{q}^{\,-}}\psi_{L}(x^i,t)\right)= q\varepsilon A\,B_L,
\end{equation}
where $\vec p_{q}$ corresponds to the vector position of the $q$ defects enclosed by the path. Consequently, the effect on the screen will be the same with a greater phase shift given by $\theta=\arctan(q\varepsilon k^2_1k_2/k^2)$. Hence, an increment in the number of defects in the lattice can be effectively described by an amplification of the depth, represented by the quantity $\varepsilon$, of one single defect.

\subsection{Wave packet solution}
Now, let us consider the effect of the torsion field on a wave packet. The wave function of a packet produced at $x^{i}_0$ satisfying the homogeneous part of Eq.\ (\ref{eq30}) is given by
\begin{equation}
\label{wave_pac}
\psi_0(x^i,t) =\frac{a}{(2\pi)^{\frac{3}{2}}}\int^{+\infty}_{-\infty}e^{-\frac{a^2}{4}(\Delta k^i)^2}\textrm{e}^{i[k_j (\Delta x^j)-\omega t]}\textrm{d}^2k\, ,
\end{equation}
where $a$ measures the width of the packet and we also defined $\Delta A^i\equiv A^i-A^i_0$, for $A^i=x^i,k^i$ and $\omega\equiv\hbar k^2/2m$. The integration of Eq.\ (\ref{wave_pac}) yields
\begin{eqnarray}
\label{pac_psi_0}
\psi_0(x^i,t) &=&\frac{2a}{\sqrt{2\pi}}\frac{\mathrm{e}^{-i\xi}}{\left(a^4 + 4\hbar^2t^2/m^2\right)^{\frac{1}{2}}}\mathrm{e}^{i(k_j)_0(\Delta x^j)}\nonumber\\
&&\times\mathrm{exp}\left[-\frac{(\Delta x^j-\frac{\hbar t}{m}k^j_0)^2}{(a^2+2i\hbar t/m)}\right]\, ,
\end{eqnarray}
with $\xi=\hbar\,t\,k^2_0/(2m)+\arctan[2\hbar\, t/(a^2m)]$.
\begin{figure}[!ht]
\includegraphics[width=0.48\textwidth]{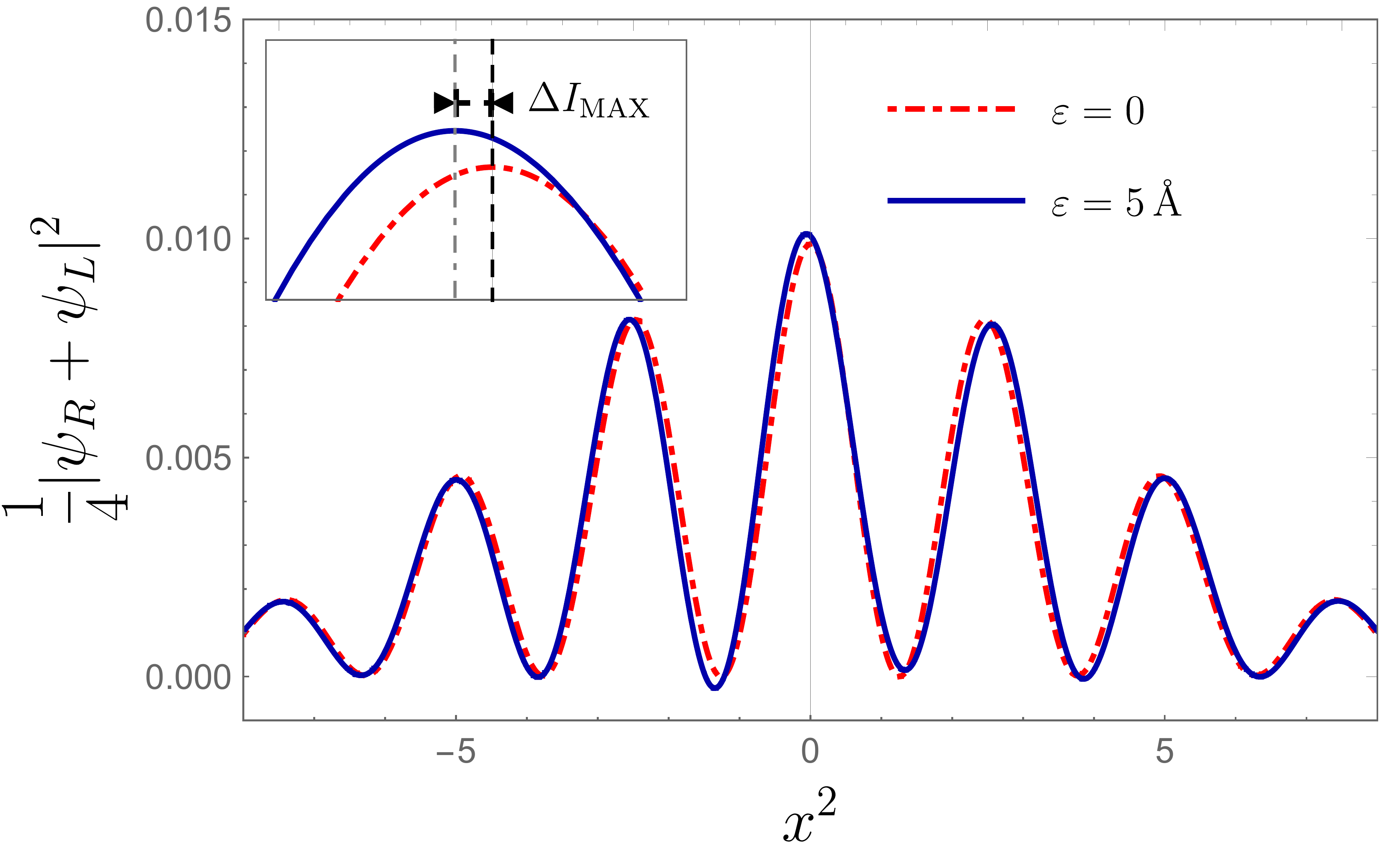}
\caption{Interference fringes pattern of a double-slit experiment in the presence of torsion generated by a screw dislocation (solid line) compared with the same setup in the absence of torsion (dot-dashed line).}
\label{fig2}
\end{figure}
Still regarding the double-slit experiment, the modification due to torsion on the left-going wave will be just the superposition of the modification on each plane wave component, as follows
\begin{eqnarray}
\label{pac_psi_1}
\psi_{1}(x^i,t)&=&-\frac{i\,a}{(2\pi)^{\frac{3}{2}}}\int^{+\infty}_{-\infty}\frac{k_1^2k_2}{k^2}\,\textrm{e}^{-\frac{a^2}{4}\Delta k_i^2}\textrm{e}^{i(k_jx^j-\omega t)}\textrm{d}^2 k\nonumber\\
&=&-iB_L(\x,t)\psi_0(\x,t)\, ,
\end{eqnarray}
with  (see Appendix \ref{appA})
\begin{eqnarray}
&&B_L(x^i,t)=\frac{\alpha}{\pi}\int\textrm{d}^2k \,\frac{k_1^2k_2}{k^2}\textrm{e}^{-\alpha(k^i-\beta^i)^2}\nonumber\\
&&=\frac{\beta_2}{\alpha^2\beta^6}\left[(\beta^2-4\beta^2_1)\left(\mathrm{e}^{-\alpha\beta^2}+\alpha\beta^2-1\right)+2\alpha^2\beta^4\beta^2_1\right],\nonumber\\
\label{eq38}
\end{eqnarray}
where we defined the auxiliary quantities
\begin{align}
\alpha=\frac{1}{4}\left(a^2+\frac{2i\hbar t}{m}\right),\hspace{5mm}
\beta_j=\frac{1}{2\alpha}\left(\frac{a^2}{2}(k_j)_0+ix_j\right),
\end{align}
and $\beta^2\equiv\beta_1^2+\beta^2_2$. Therefore, if the right-going and left-going waves travel through the paths $x^i_{(1)}$ and $x^i_{(2)}$ previously defined, the squared amplitude of the wave packet on the detection plate will be
\begin{eqnarray}
|\Psi|^2\sqrt{g}\,\mathrm{d}^2x\approx\frac{1}{4}|\psi_R(x^i_{(1)})+\psi_L(x^i_{(2)})|^2\nonumber\\
=\frac{1}{4}|\psi_0(x^i_{(1)})+\psi_0(x^i_{(2)})|^2\nonumber\\
-\frac{\varepsilon}{2}\mathrm{Im}\left[\psi^*_0(x^i_{(1)})\psi_0(x^i_{(2)})B_L(x^i_{(2)})\right].\label{psi2_ds_t}
\end{eqnarray}

For the sake of illustration, we compare the interference fringes pattern provided by Eq.\ (\ref{psi2_ds_t}) with the one obtained from the same setup but without torsion. The difference between them is depicted in Fig.\ (\ref{fig2}), where we clearly see the break of symmetry due to torsion and a tiny deviation in the size of the first peaks. There, the free parameters were chosen as follows: the slits are separated by $10 \,\mathrm{nm}$, their apertures are $0.1\,\mathrm{nm}$ each (which coincide with the wave packet width $a$), and  the screen of detection is at $20\,\mathrm{nm}$ from the plane $x^1=0$. Also, $(k_1)_0=50 \,\mathrm{nm}^{-1}$, $(k_2)_0=0$ and the instant time as $t=m x^1 /(k_1)_0 \hbar$, such that the peak of the packet in the $x^1$-direction is depicted. Thus, the displacement of the first peaks are of the order $\Delta I_{MAX}\sim 10^{-2}\,\mathrm{nm}$, corresponding to a detectable deviation at the atomic scale of $0.1$ angstroms. This first order approach breaks down at the valleys, where  higher orders of perturbation in $\varepsilon$ are required in order to guarantee the positive-definiteness of the ordenate axis. Nevertheless, the existence of a shift in the interference pattern is preserved at the length scale aforementioned.

The double-slit is one among many other experiments that can probe interference effects at such scale, for instance, the detection of the Aharonov-Bohm shift using carbon nanotubes \cite{bachtold} or phase measurements in quantum dots \cite{schuster}. Moreover, as the defect contributes solely for closed paths, experiments done with one slit closed at a time \cite{sawant2014,sinha2010}, in addition to the two slits, could be useful to single out the interference contribution. 

Note that the possibility of an experimental test of the small correction predicted for the interference fringes pattern is a consequence of the deltalike profile of torsion, breaking the spatial symmetry of the phenomenon with respect to the torsion-free case and slightly changing the amplitude and the position of the peaks.

The assumption of taking the lattice as a continuous surface is fulfilled as long as the Compton wavelength $\lambda_c=h/mc$ is much greater than the lattice spacing. This is not a problem, since the mass of the particle does not change the interference pattern, but only the time scale of the experiment.

\section{Attempt to save the probabilistic interpretation}
Apart from the nonunitarity, the major drawback of Schr\"odinger equation in the presence of torsion corresponds to the violation of Born's rule, which states: \textit{the only real physical information is given by the modulus squared of the wave function amplitude, that is, the normalized and conserved probability density defined all over the space}. This kind of trouble also appears in effective quantum open systems, where non-Hermitian Hamiltonians are well suited to treat dissipative localized systems embedded in an environment. This is the case of the so-called \textit{quantum thermodynamics} \cite{kosloff,hu2018}, where the system of interest is part of a large closed system governed by an unitary evolution with interaction between the local system and the bath encoded in a memory kernel (in general, the evolution of the small system is driven by an integro-differential equation). Another example of non-Hermitian quantum theory is the one proposed in Ref.\ \cite{bender2007}, for which the quantum description is performed by postulating a spacetime reversal ($\mathcal{PT}$) symmetric Hamiltonian instead of an hermitian one. It is possible to demonstrate that the spectrum of such $\mathcal{PT}$-invariant Hamiltonians is still real and positive, but the inner product should be modified in order to preserve the new symmetry. Note that it is different from Ref.\ \cite{kleinert2000} where the usual inner product is altered by adopting a weight function to make the Laplacian a hermitian operator, that can be attained when one has a scalar torsion. In a more fundamental description, such problems can be circumvented by other formulations of quantum mechanics, where the probabilistic interpretation is secondary, such as the De Broglie-Bohm theory \cite{bohm52a,bohm52b}.

Notwithstanding, the deltalike torsion investigated here does not change the energy levels, as one can see by the solutions given above. Therefore, it is not possible to interpret our quantum system in terms of dissipative processes. We, thus, propose a solution to this issue by appropriately redefining the probability distribution in a way that it can also be applied to other systems whose corrections affect only sets of measure zero. %

In order to effectively see the nonunitary nature of the problem, we must calculate the continuity equation determining the flow of probability of the system. Using the modified Schr\"odinger equation (\ref{eq30}), we find that
\begin{equation}
\begin{array}{l}
\displaystyle\fracc{\partial}{\partial t}\int|\Psi|^2\sqrt{g}\,\rmd^2x
\approx \int\fracc{\partial}{\partial t}|\psi|^2\rmd^2x\\[1ex]
=-\displaystyle\int\left[\partial_k j^k+\frac{\varepsilon\hbar}{m}\delta(x^i)\textrm{Im}(\psi^*\partial_1\psi)\right]\rmd^2x\, ,
\end{array}
\end{equation}
with the usual definition of a free particle current density $j_k=(\hbar/m)\textrm{Im}(\psi^*\partial_k\psi)$. Thus,
\begin{equation}
\label{cont-eq1}
    \int\left(\frac{\partial}{\partial t}|\psi|^2+\partial_k j^k\right)\rmd^2\x=-\frac{\varepsilon\hbar}{m}\textrm{Im}(\psi_0^*\partial_1\psi_0)|_{\x=0}.
\end{equation}
On the left-hand side, we have the continuity equation terms as the integrand, but the right-hand side is not zero whenever there is torsion ($\varepsilon\neq0$). 

Now, the wave functions discussed before can be written down as
\begin{equation}\label{disc-func1}
\psi(x^i,t) = \begin{cases}
             \psi_0(x^i,t)  & \text{if } x^1 < 0\, , \\
             \psi_0(x^i,t)\left[1-\frac{i\varepsilon}{2}B_L(x^i,t)\right]  & \text{if } x^1 > 0\, ,
       \end{cases} \quad
\end{equation}
for which the modulus squared is trivial when $x^1<0$, while for $x^1>0$, to first order in $\varepsilon$, we have
\begin{equation}
    |\psi(x^i,t)|^2\approx|\psi_0|^2\left(1+\varepsilon\textrm{Im}B_L\right)\, .
\end{equation}
For a Gaussian wave packet produced in the negative $x^1$-axis away from the defect, and initially normalized, the source term of Eq.\ (\ref{cont-eq1}) reads
\begin{multline}
\textrm{Im}(\psi_0^*\partial_1\psi_0)|_{\x=0}\\
  =\frac{2a^2\left(a^4k^1_{0}-\frac{4\hbar\,t}{m}x^1_0\right)}{\pi(a^4+\frac{4\hbar^2t^2}{m^2})^2}\textrm{exp}\left[-2a^2\frac{\left(x^i_0+\frac{\hbar t}{m}k^i_0\right)^2}{a^4+\frac{4\hbar^2t^2}{m^2}}\right].
\end{multline}
This is always positive for the wave function given by Eqs. (\ref{pac_psi_0})-(\ref{pac_psi_1}), ensuring a draining of probability in the defect. However, if we reflect the problem with respect to the $x^2$-axis, so that $k_0<0$ and $x_0>0$, the system gains probability from the dislocation. 

Indeed, Eq.\ (\ref{cont-eq1}) poses some difficulties for the usual probabilistic interpretation of the wave function. One can imagine a very localized system to be detected in a region comprising it, so that the probability should be $1$. If there is a defect in this region, the detector could never click or could have a probability to click greater than one without any change in its energy.

It is possible to circumvent this issue by modifying the Born rule accordingly. In fact, if we extend the integration region to be large enough, we can discard the term
$\partial_k j^k$. Even for the discontinuous wave function (\ref{disc-func1}), one can verify that this term vanishes by dividing the integration region, using the divergence theorem twice and the fact that we are dealing with square-integrable functions. By doing this, we realize that we can preserve a probabilistic interpretation by redefining the probability density function as
\begin{eqnarray}
    \widetilde{{\cal P}}(x^i,t)={\cal P}(x^i,t)+\frac{\varepsilon\hbar}{m}\delta(x^i)\int_{t_0}^t \textrm{Im}(\psi^*\partial_1\psi) dt\, ,
\end{eqnarray}
where $t_0$ is the instant of time that normalizes the new probability distribution as $\int_{\Omega} \bar{{\cal P}}d^2x=1$, for an arbitrarily large integration region $\Omega$. This guarantees that somewhere in space the particle described by the wave function must be detected.

The expectation value of a given function of the spatial coordinates can now be defined as $\langle V(x^i,t)\rangle =\int \sqrt{g}\bar{{\cal P}}V(x^i,t) d^2x$. Notice that actual departures from the usual Born rule shall emerge when assuming integration regions that comprise the location of the defect. However, the probability of locating the particle perfectly at the origin is not well defined. This is not a limitation of this approach, since, for practical reasons, one shall always consider an extended region for integration. The expectation value of quantities that depend on momenta can be calculated through the Fourier transform of the wave function and those quantities whose dependence involves both momentum and spatial coordinate, it shall be done by a combination thereof. 

Ultimately, we also find the inner product between states by using the polarization identity\footnote{Given a normed vector space $({\cal H},||.||)$ over the complex numbers, the inner product (antilinear in the first argument) between two vectors $x\, , y \in {\cal H}$ is given by
\begin{equation}
    \langle x|y\rangle=\frac{1}{4}(||x+y||^2-||x-y||^2-i||x+iy||^2+i||x-iy||^2)\, .
\end{equation}} (see Proposition $14.1.12$ of \cite{book-blachard})
that allows one to derive it from a given norm, which in our case is defined by the modified Born rule $||\Psi||^2=\widetilde{\langle\Psi|\Psi\rangle}=\int d^2x\sqrt{g}\widetilde{{\cal P}}$. This way, the modified inner product involving wave functions $\Psi_{1,2}$ is found to be
\begin{align}
  &\widetilde{\langle\Psi_1|\Psi_2\rangle} = \int d^2x\sqrt{g}\Psi_1^{\ast}\Psi_2\\ &+\int d^2x\left[ \frac{\epsilon \hbar}{m}\delta(x^i)\int_{t_0}^{t} dt\left(\frac{\psi_1^{\ast}\partial_1\psi_2-\psi_2\partial_1\psi_1^{\ast}}{2i}\right)\right]\, .\nonumber
\end{align}
As one can see, this expression reduces to the modified Born rule when $\Psi_1=\Psi_2$. This statement is in agreement with comments made in section $11$ of \cite{kleinert2000}, where it is stressed that, for arbitrary torsion fields, the appropriate inner product that makes Hermitian the torsionful Laplacian operator should be a matter of investigation for each case -- unless one has a gradient torsion, such that the new inner product is straightforwardly found \cite{kleinert2000}, which is not the case of the present paper.

\section{Final Remarks}
The aforementioned statements are not exclusive to the model of torsion developed here, but it can be applied to any quantum system driven by the Schr\"odinger equation with topological defects modeled by a localized torsion field. As long as the wave packet admits a region where the torsion is negligible, the interaction between them can be described in the same lines as a scattering process, which is valid for both plane waves and wave packets. 

We have also shown here that the small correction predicted by our results to the interference fringes pattern due to the deltalike torsion could be measured by feasible experimental tests. Those modifications in the interference phenomena are expected to be general, as the presence of torsion is introduced through a nonholonomical mapping, breaking the spatial symmetry of the wave function propagation.

The probabilistic interpretation can be adapted to this effective and perturbative situation, since the deviation from the torsion-free case is very small and such smallness may be experimentally measured. Furthermore, the redefinition of the probability distribution used to calculate the expected value of observables guarantee the control of such discrepancy along time. 

\section*{Acknowledgments}
We are in debts with the anonymous Referee for the valuable comments on a previous version of this paper. I. P. L. would like to acknowledge the contribution of the COST Action CA18108 and was partially supported by the National Council for Scientific and Technological Development - CNPq grant 306414/2020-1. A. L. F. J. was supported by the Research Support Foundation of Esp\'irito Santo - FAPES grant number 13/2019.
\\

\appendix

\section{Calculating the torsion contribution $B_L(\x,t)$}\label{appA}

In order to solve the integral in Eq.(\ref{eq38}), we first change for spherical coordinates in k-space, so that
\begin{multline}
    B_L(\x,t)=\frac{\alpha}{\pi}\int\textrm{d}^2\vec k \,\frac{k_1^2k_2}{k^2}\textrm{e}^{-\alpha(\vec k-\vec \beta)^2}=\frac{\alpha}{\pi}\mathrm{e}^{-\alpha\beta^2}\\\times\int_0^{\infty}\mathrm{d}r\, r^2\mathrm{e}^{-\alpha r^2}
    \int_0^{2\pi}d\theta \,\mathrm{cos}^2\theta\,\mathrm{sin}\theta\,\mathrm{e}^{2\alpha\,r(\beta_1\mathrm{cos}\theta+\beta_2\textrm{sin}\theta)}.
    \label{eqa1}
\end{multline}
Then, using the identity \cite{grad07}
\begin{align}
    &\int_0^{2\pi}d\theta \,\mathrm{cos}^2\theta\,\mathrm{sin}\theta\,\mathrm{e}^{a\mathrm{cos}\theta+b\textrm{sin}\theta}\nonumber\\
    &=\frac{\partial^2}{\partial a^2}\frac{\partial}{\partial b}\int_0^{2\pi}d\theta \,\mathrm{e}^{a\mathrm{cos}\theta+b\textrm{sin}\theta}=2\pi\frac{\partial^2}{\partial a^2}\frac{\partial}{\partial b}I_0\left(\sqrt{a^2+b^2}\right)\nonumber\\
    &=2\pi b\frac{I_2\left(\sqrt{a^2+b^2}\right)}{\left(\sqrt{a^2+b^2}\right)^2}+2\pi a^2b \frac{I_3\left(\sqrt{a^2+b^2}\right)}{\left(\sqrt{a^2+b^2}\right)^3},
\end{align}
where $I_\nu(z)$ is the modified Bessel function of the first kind, back into Eq. (\ref{eqa1}), it gives
\begin{align}
\label{bl_int}
    B_L(\x,t)&=2\alpha\beta_2\mathrm{e}^{-\alpha\beta^2}\left[\frac{1}{2\alpha\beta^2}\int_0^{\infty}\mathrm{d}r\, r\mathrm{e}^{-\alpha r^2}I_2\left(2\alpha\,r\sqrt{\beta^2}\right)\right.\nonumber\\
    &\left.+\frac{\beta^2_1}{\beta^3}\int_0^{\infty}\mathrm{d}r\, r^2\mathrm{e}^{-\alpha r^2}I_3\left(2\alpha\,r\sqrt{\beta^2}\right)\right].
\end{align}
Now, using the series expansion of $I_\nu(z)$:
\begin{equation}
    I_\nu(z)=\sum^{\infty}_{n=0}\frac{(z/2)^{\nu+2n}}{\Gamma(n+1)\Gamma(n+\nu+1)},
\end{equation}
we can simplify the integrals in Eq.\ (\ref{bl_int}), as follows:
\begin{align}
   & \int_0^{\infty}\mathrm{d}r\, r\mathrm{e}^{-\alpha r^2}I_2\left(2\alpha\,r\sqrt{\beta^2}\right)\nonumber\\
    &=\sum^{\infty}_{n=0}\frac{(\alpha\sqrt{\beta^2})^{2+2n}}{\Gamma(n+1)\Gamma(n+3)}\int_0^{\infty}\mathrm{d}r\, r^{3+2n}\mathrm{e}^{-\alpha r^2}\nonumber\\
    &=\frac{1}{2\alpha}\sum^{\infty}_{n=0}\frac{\Gamma(n+2)}{\Gamma(n+1)\Gamma(n+3)}(\alpha^{\frac{1}{2}}\sqrt{\beta^2})^{2+2n}\nonumber\\
    &=\frac{1}{2\alpha^2\beta^2}\left[1+(\alpha\beta^2-1)\mathrm{e}^{\alpha\beta^2}\right],
\end{align}
and
\begin{align}
   & \int_0^{\infty}\mathrm{d}r\, r^2\mathrm{e}^{-\alpha r^2}I_3\left(2\alpha\,r\sqrt{\beta^2}\right)\nonumber\\
   &=\sum^{\infty}_{n=0}\frac{(\alpha\sqrt{\beta^2})^{3+2n}}{\Gamma(n+1)\Gamma(n+4)}\int_0^{\infty}\mathrm{d}r\, r^{5+2n}\mathrm{e}^{-\alpha r^2}\nonumber\\
    &=\frac{1}{2\alpha^{\frac{3}{2}}}\sum^{\infty}_{n=0}\frac{\Gamma(n+3)}{\Gamma(n+1)\Gamma(n+4)}(\alpha^{\frac{1}{2}}\sqrt{\beta^2})^{3+2n}\nonumber\\
    &=\frac{1}{2\alpha^3(\beta^2)^{3/2}}\left[-2+(2-2\alpha\beta^2+\alpha^2\beta^4)\mathrm{e}^{\alpha\beta^2}\right].
\end{align}

\end{document}